# Quadratic Hierarchy Flavor Rule as the Origin of Dirac CP-Violating Phases

## E. M. Lipmanov

40 Wallingford Road # 272, Brighton MA 02135, USA

### Abstract

The premise of an organizing quadratic hierarchy rule in lepton-quark flavor physics was used for unified description of hierarchy patterns of four generic pairs of flavor quantities 1) charged-lepton and 2) neutrino deviations from mass-degeneracy, 3) deviations of lepton mixing from maximal magnitude and 4) deviations of quark mixing from minimal one. Here it is shown that that quadratic hierarchy equation that is uniquely related to three flavor particle generations may have yet another important function. It determines two complementary values of the Dirac CP-violating phases in the quark $\delta_q \cong 58.8°$ and neutrino $\delta_{nu} \cong 31.2°$ mixing matrixes without involvement of any empirical parameters. If supported by accurate experimental data, that quadratic hierarchy rule is an explicit CP-violation source in elementary particle mixing matrixes and a single source of CP-violation at least in the quark mixing matrix phenomenology.

## 1. Introduction

One of the great unsolved problems in flavor elementary particle physics is still the empirical CP-violation fact. It calls for three flavor generations. The Dirac CP-violating phase in the phenomenological mixing matrix of three quark generations was first introduced in ref. [1].

In ref. [2,3] I considered a presumably basic universal quadratic hierarchy rule for deviation-from-mass-degeneracy (DMD) flavor physical DMD-quantities that also calls for three elementary particle generations and determines the *patterns* of most known dimensionless flavor quantities such as charged lepton (CL) and quasi-degenerate (QD) neutrino mass ratios and large neutrino and small quark mixing angles and their complementarity relations. The absolute values of all these quantities are expressed through one empirical universal physical parameter $\alpha_o \cong e^{-5}$. In this publication I show that the same quadratic flavor hierarchy rule leads to a suggested algebraic equation for the Dirac phase $\delta$ with two basic solutions independent of any empirical parameters. One of these $\delta$-values lies well within known experimental ranges of the quark CP-violating phase and should be the accurate value of this phase, while the only other (complementary value) should be the unknown Dirac CP-violating phase of the neutrino mixing matrix. These results suggest quantitative unification of the Dirac CP-violating phases in the mixing matrixes of quarks and leptons in addition to the unification of the quark and neutrino mixing angles.

## 2. Basic quadratic hierarchy equation in flavor physics

In ref. [2,3] I considered the universal quadratic hierarchy equation for flavor physical DMD-quantities,

$$[DMD(2)]^2 \cong 2[DMD(1)], \qquad (1)$$

where DMD(n), n=1,2, denote deviations from unity of the relevant lepton flavor dimensionless quantities: 1) particle mass ratio squared and 2) particle mixing parameters. The hierarchy rule (1) answers the specific quantitative neutrino-quark problem of two empirically large solar and atmospheric mixing parameters $Sin^2 2\theta_{12}$ and $Sin^2 2\theta_{23}$ and its relation to two small quark mixing parameters. The particular hierarchy equations for neutrino and quark mixing parameters are given [3] by

$$(Sin^2 2\theta_{12} -1)^2 \cong 2 \left| Sin^2 2\theta_{23} -1 \right|,$$
$$(Cos^2 2\theta_c - 1)^2 \cong 2 \left| Cos^2 2\theta' - 1 \right|, \qquad (2)$$

where $\theta_c$ is the Cabibbo angle and $\theta'$ is the next to the largest quark mixing angle. The DMD(n)-quantities from (1) are interpreted in (2) as *deviations* from maximal or minimal mixing for neutrinos or quarks respectively.

Four interesting solutions of the hierarchy Eq.(1) with dual, large and small, extended DMD-quantities are considered [3]. The solution of Eq.(1) for large DMD-quantities is for CL-mass ratios[1]

$$(m_\tau/m_\mu)^2 \cong 2/\alpha_o, \quad (m_\mu/m_e)^2 \cong 2/\alpha_o^2, \quad \alpha_o = e^{-5}. \qquad (3)$$

Three solutions of Eq.(1) are obtained for small DMD-quantities. 1) QD-neutrino mass ratios

$$(m_3^2/m_2^2) \cong \exp(2r), \quad (m_2^2/m_1^2) \cong \exp(2r^2), \qquad (4)$$

---

[1] The quark mass ratios are not considered explicitly.

where $(m_1 < m_2 < m_3)$ are the neutrino masses and $r$ is the neutrino oscillation solar-atmospheric hierarchy parameter $r = (m_2{}^2 - m_1{}^2)/(m_3{}^2 - m_2{}^2) \cong 5\alpha_o$.

2) Small deviations of large neutrino mixing parameters from maximal mixing

$$\text{Sin}^2\,2\theta_{12} \cong \exp(-2\sqrt{\alpha_o}) \cong 0.8486,\ \theta_{12} \cong 33.6^\circ,$$

$$\text{Sin}^2\,2\theta_{23} \cong \exp(-2\alpha_o) \cong 0.9866,\ \theta_{23} \cong 41.7^\circ. \qquad (5)$$

3) Small deviations of quark mixing parameters from minimal mixing

$$\text{Cos}^2\,2\theta_c \cong \exp(-2\sqrt{\alpha_o}) \cong 0.8486,\ 2\theta_c \cong 22.9^\circ,$$

$$\text{Cos}^2\,2\theta' \cong \exp(-2\alpha_o) \cong 0.9866,\ 2\theta' \cong 6.6^\circ. \qquad (6)$$

For illustration of the duality-relation between large and small DMD-solutions in (3)-(6) consider the virtual limit $\alpha_o \to 0$: the divergence of CL masses gets infinitely large and the CL mixing (if present) disappears, the neutrinos get exactly mass-degenerate and neutrino mixing is getting maximal, the divergences of quark mass spectra are getting infinitely large and (naturally) the quark mixing disappears.

## 3. Quadratic hierarchy flavor rule as origin of Dirac CP-violating phases

Dirac CP-violating phases enter the mixing matrixes of quarks $V_{CKM}$ and neutrinos $V_{nu}$ in the form of the exponential factor $e^{i\delta}$. As of to date, there is no known connection between the two Dirac phases for quarks $\delta_q$ and neutrinos $\delta_{nu}$.

The real and imaginary parts, $\text{Cos}\delta$ and $\text{Sin}\delta$, of the exponential factor $e^{i\delta}$ are a pair of two generic flavor quantities related to three particle generations in the sense of ref.[3] and should obey the hierarchy equation (1) for generic flavor pairs. Since in this peculiar case both of the generic quantities depend on the same unknown $\delta$-parameter, the hierarchy Eq.(1) should determine these parameters uniquely.

The physical meaning of Eq.(1) for the $\delta$-angles is that it defines a quadratic relation between the deviations from maximal and minimal CP-violating phase values $(1-\mathrm{Sin}^2\delta)$ and $(1-\mathrm{Cos}^2\delta)$ respectively. Since its solution is symmetric under exchanges of $\mathrm{Sin}^2\delta$ and $\mathrm{Cos}^2\delta$, the two solutions must be complementary

$$\delta_1 + \delta_2 = \pi/2. \qquad (7)$$

Consider two sole possibilities:

1). $\mathrm{Sin}^2\delta_1 > \mathrm{Cos}^2\delta_1$. In that case, $\mathrm{DMD}(2)=(1 - \mathrm{Cos}^2\delta_1)$, $\mathrm{DMD}(1)=(1 - \mathrm{Sin}^2\delta_1)$, and from (1) follows a definite equation for the phase $\delta_1$,

$$(1 - \mathrm{Cos}^2\delta_1)^2 \cong 2(1 - \mathrm{Sin}^2\delta_1), \qquad (8)$$

or

$$\mathrm{Sin}^4\delta_1 \cong 2\,\mathrm{Cos}^2\delta_1. \qquad (8')$$

Its only solution is

$$\mathrm{Cos}^2\delta_1 \cong 2 - \sqrt{3} \cong 0.268, \ \delta_1 \cong 58.8^{\circ}. \qquad (9)$$

2). $\mathrm{Sin}^2\delta_2 < \mathrm{Cos}^2\delta_2$, $\mathrm{DMD}(2) \leftrightarrow \mathrm{DMD}(1)$, and the equation for $\delta_2$ is given by

$$\mathrm{Cos}^4\delta_2 \cong 2\,\mathrm{Sin}^2\delta_2, \qquad (10)$$

with the solution

$$\mathrm{Sin}^2\delta_2 \cong 0.268, \ \delta_2 \cong 31.2^{\circ}. \qquad (11)$$

Those two values for the Dirac phases $\delta$ are the only solutions of the hierarchy equation (1) independent of any empirical parameters.

The first solution $\delta_1 \cong 58.8°$ is well within the experimental data ranges of the CP-violating phase of the quark CKM-mixing matrix [4], $\delta_{CKM} = 59° \pm 13°$. So, the obtained value $\delta_1$ is the prediction for the quark CP-violating phase

$$\delta_q \cong 58.8°. \qquad (12)$$

Since the two values for the Dirac phase are uniquely connected by the universal hierarchy rule, it is natural to identify the second solution $\delta_2 \cong 31.2°$ with the unknown value of the CP-violating Dirac phase $\delta_{nu}$ of the neutrino mixing matrix

$$\delta_{nu} \cong 31.2°. \qquad (13)$$

The value (13) is the predicted Dirac CP-violating phase in the neutrino mixing matrix.

To summarize, the quadratic hierarchy equation as a rule in flavor physics may be the origin of the Dirac phases in quark and lepton mixing matrixes and predicts: 1) necessary CP-violation in the quark mixing matrix – not vanishing Dirac phase $\delta_q \neq 0$, 2) definite and not maximal or minimal values (12) and (13) for the Dirac CP-violating phases in the united quark and lepton mixing matrixes, 3) the pair of quark and lepton Dirac phase values is fully determined by the hierarchy rule (1) in contrast to the pairs of mixing angle values related to the parameter $\alpha_o$, 4) quark-lepton complementarity is a common feature of the mixing angles and Dirac CP-violating phases[2].

---

[2] The possibility of quark-lepton complementarity of the Dirac CP-violating phases was mentioned in [5].

## 4. Conclusion

In accordance with the above results, the quadratic hierarchy rule (1) should be one of the main regularities in low energy elementary particle flavor physics and the origin of CP-violation at least in the quark mixing matrix phenomenology.

1. The hierarchy rule is unique only for three particle flavor generations. The Dirac phases of quarks and leptons $\delta_q$ and $\delta_{nu}$ are also unique only in case of three generations. That common unique relation to three particle generations resulted in the basic connection between CP-violation and the quadratic hierarchy rule in flavor physics and leads to quantitative predictions of the Dirac CP-violating phases (12) and (13).

2. It is the source of two Dirac CP-violating phases, and probably the only source of the empirically known CP-symmetry violation at least in the *quark* mixing matrix.

3. It suggests complementarity relations of the Dirac CP-violating phases for quarks $\delta_q \cong 58.8°$ and leptons $\delta_{nu} \cong 31.2°$ analogous to such relations between lepton and quark doubled mixing angles.

4. By analogy with the known simple and elegant regularities of classical low energy mechanics and low energy quantum mechanics, the *quadratic hierarchy rule* (1) may be the basic approximate law of CP-noninvariant low energy nonrelativistic flavor physics.

5. By its inferences, hierarchy rule (1) means a new quantitative approach to standard model mixing matrix phenomenology in terms of a universal rule for mass-symmetry-violating extended DMD-quantities not relied on a particular type of exact flavor symmetry.